\documentclass[useAMS,usenatbib]{mn2e}

\usepackage{graphicx}
\usepackage{amsmath}
\usepackage{amssymb}
\usepackage{dcolumn}

\usepackage{longtable}

\usepackage{color}
\usepackage{soul}
\usepackage{ulem}

\definecolor{darkblue}{rgb}{0.0, 0.0, 0.60}

\def\gtrsim{\mathrel{\hbox{\rlap{\hbox{\lower4pt\hbox{$\sim$}}}\hbox{$>$}}}}
\def\ltsim{\mathrel{\hbox{\rlap{\hbox{\lower4pt\hbox{$\sim$}}}\hbox{$<$}}}}

\usepackage{hyperref}
\hypersetup{bookmarks=true, colorlinks=true, linkcolor=darkblue, citecolor=darkblue, filecolor=darkblue, urlcolor=darkblue}

\title[Magnetic massive stars as progenitors of heavy BHs]{Magnetic massive stars as progenitors of ``heavy" stellar-mass black holes}

\author[Petit et al.]{Petit, V. $^{1}$\thanks{VPetit@fit.edu}, Keszthelyi, Z.$^{2,3}$, MacInnis, R.$^{1}$, Cohen, D. H.$^4$, Townsend, R. H. D.$^{5}$, \and Wade, G. A.$^2$, Thomas, S. L.$^{1}$, Owocki, S. P.$^6$, Puls, J.$^7$, ud-Doula, A.$^8$
\\
$^{1}$ Department of Physics and Space Sciences, Florida Institute of Technology, 150 W. University Blvd, Melbourne, FL 32904 \\
$^{2}$ Department of Physics, Royal Military College of Canada, PO Box 17000 Station Forces, Kingston, ON, Canada K7K 0C6\\
$^{3}$ Department of Physics, Engineering Physics and Astronomy, Queen's University, 99 University Avenue, Kingston, ON K7L 3N6, Canada\\
$^{4}$ Department of Physics and Astronomy, Swarthmore College, Swarthmore, PA 19081, USA\\
$^{5}$ Department of Astronomy, University of Wisconsin-Madison, 5534 Sterling Hall, 475 N Charter Street, Madison, WI 53706, USA\\
$^{6}$ Department of Physics and Astronomy, University of Delaware, Newark, DE, 19711, USA\\
$^{7}$  LMU Munich, Universit\"ats-Sternwarte, Scheinerstr. 1, 81679, M\"unchen, Germany\\
$^{8}$ Penn State Worthington Scranton, Dunmore, PA 18512, USA
}

\begin{document}
%
%
%


\def\jnl@style{\it}
\def\aaref@jnl#1{{\jnl@style#1}}

\def\aaref@jnl#1{{\jnl@style#1}}

\def\aj{\aaref@jnl{AJ}}                   
\def\araa{\aaref@jnl{ARA\&A}}             
\def\apj{\aaref@jnl{ApJ}}                 
\def\apjl{\aaref@jnl{ApJ}}                
\def\apjs{\aaref@jnl{ApJS}}               
\def\ao{\aaref@jnl{Appl.~Opt.}}           
\def\apss{\aaref@jnl{Ap\&SS}}             
\def\aap{\aaref@jnl{A\&A}}                
\def\aapr{\aaref@jnl{A\&A~Rev.}}          
\def\aaps{\aaref@jnl{A\&AS}}              
\def\azh{\aaref@jnl{AZh}}                 
\def\baas{\aaref@jnl{BAAS}}               
\def\jrasc{\aaref@jnl{JRASC}}             
\def\memras{\aaref@jnl{MmRAS}}            
\def\mnras{\aaref@jnl{MNRAS}}             
\def\pra{\aaref@jnl{Phys.~Rev.~A}}        
\def\prb{\aaref@jnl{Phys.~Rev.~B}}        
\def\prc{\aaref@jnl{Phys.~Rev.~C}}        
\def\prd{\aaref@jnl{Phys.~Rev.~D}}        
\def\pre{\aaref@jnl{Phys.~Rev.~E}}        
\def\prl{\aaref@jnl{Phys.~Rev.~Lett.}}    
\def\pasp{\aaref@jnl{PASP}}               
\def\pasj{\aaref@jnl{PASJ}}               
\def\qjras{\aaref@jnl{QJRAS}}             
\def\skytel{\aaref@jnl{S\&T}}             
\def\solphys{\aaref@jnl{Sol.~Phys.}}      
\def\sovast{\aaref@jnl{Soviet~Ast.}}      
\def\ssr{\aaref@jnl{Space~Sci.~Rev.}}     
\def\zap{\aaref@jnl{ZAp}}                 
\def\nat{\aaref@jnl{Nature}}              
\def\iaucirc{\aaref@jnl{IAU~Circ.}}       
\def\aplett{\aaref@jnl{Astrophys.~Lett.}} 
\def\apspr{\aaref@jnl{Astrophys.~Space~Phys.~Res.}}
\def\bain{\aaref@jnl{Bull.~Astron.~Inst.~Netherlands}} 
\def\fcp{\aaref@jnl{Fund.~Cosmic~Phys.}}  
\def\gca{\aaref@jnl{Geochim.~Cosmochim.~Acta}}   
\def\grl{\aaref@jnl{Geophys.~Res.~Lett.}} 
\def\jcp{\aaref@jnl{J.~Chem.~Phys.}}      
\def\jgr{\aaref@jnl{J.~Geophys.~Res.}}    
\def\jqsrt{\aaref@jnl{J.~Quant.~Spec.~Radiat.~Transf.}}
\def\memsai{\aaref@jnl{Mem.~Soc.~Astron.~Italiana}}
\def\nphysa{\aaref@jnl{Nucl.~Phys.~A}}   
\def\physrep{\aaref@jnl{Phys.~Rep.}}   
\def\physscr{\aaref@jnl{Phys.~Scr}}   
\def\planss{\aaref@jnl{Planet.~Space~Sci.}}   
\def\procspie{\aaref@jnl{Proc.~SPIE}}   

\let\astap=\aap
\let\apjlett=\apjl
\let\apjsupp=\apjs
\let\applopt=\ao

\date{
Accepted  ?. Received 2016 October 24; in original form 2016 October 24}
\pagerange{\pageref{firstpage}--\pageref{lastpage}} \pubyear{2010}
\maketitle
\label{firstpage}

\begin{abstract}

The groundbreaking detection of gravitational waves produced by the inspiralling and coalescence of the black hole (BH) binary GW150914 confirms the existence of ``heavy" stellar-mass BHs with masses $>25\,M_\odot$. Initial modelling of the system by \citet{2016ApJ...818L..22A} supposes that the formation of black holes with such large masses from the evolution of single massive stars is only feasible if the wind mass-loss rates of the progenitors were greatly reduced relative to the mass-loss rates of massive stars in the Galaxy, concluding that heavy BHs must form in low-metallicity ($Z\ltsim 0.25-0.5\ \,Z_\odot$) environments. 
However, strong surface magnetic fields also provide a powerful mechanism for modifying mass loss and rotation of massive stars, independent of environmental metallicity \citep{2002ApJ...576..413U,2008MNRAS.385...97U}. 
In this paper we explore the hypothesis that some heavy BHs, with masses $>25\,M_\odot$ such as those inferred to compose GW150914, could be the natural end-point of evolution of magnetic massive stars in a solar-metallicity environment.
Using the MESA code, we developed a new grid of single, non-rotating, solar metallicity evolutionary models for initial ZAMS masses from 40-80 $\,M_\odot$ that include, for the first time, the quenching of the mass loss due to a realistic dipolar surface magnetic field.
The new models predict TAMS masses that are significantly greater than those from equivalent non-magnetic models, reducing the total mass lost by a strongly magnetized 80 $\,M_\odot$ star during its main sequence evolution by 20 $\,M_\odot$. This corresponds approximately to the mass loss reduction expected from an environment with metallicity $Z=1/30\,Z_\odot$.

\end{abstract}


\section{Introduction}

On Sept. 24, 2015, the Laser Interferometric Gravitational-wave Observatory (LIGO) detected their first gravitational wave event  GW150914, as predicted by \citet{2016A&A...588A..50M}. According to  \citet{2016PhRvL.116f1102A}, this event originated from the merger of two black holes with mass of 36$\,M_\odot$ and 29$\,M_\odot$  at redshift $z=0.09$. 
The high masses of the merging black holes is in stark contrast with the handful of black holes ($<15 \,M_\odot$) in our Galaxy for which dynamical masses can be inferred \citep[e.g.][]{2010ApJ...725.1918O}.
GW150914 therefore provides the best evidence that relatively ``heavy'' ($>25\,M_\odot$) black holes do form in nature. 

The most likely origin of these objects is via the evolution of massive stars. According to standard narratives of stellar evolution, one of the critical aspects to the formation of heavy black holes (in isolation as well as in multiple systems) is the total mass lost during their evolution, which in turn is very dependent on the metallicity. 

This is because massive stars have powerful, radiatively-driven stellar winds \citep[][and references therein]{2008A&ARv..16..209P} with the opacity of resonance-line transitions in the UV as the main driving mechanism. The predicted slightly sub-linear dependence of mass-loss rate on metallicity  \citep{2001A&A...369..574V} is corroborated by observations of massive stars in nearby, metal-poor galaxies \citep{2007A&A...465.1003M,2007A&A...473..603M}.  
Models of isolated, single massive star evolution show that heavy black holes are likely to form in low metallicity environments with $Z\ltsim 0.1 \,Z_\odot$ \citep{2015MNRAS.451.4086S, 2016Natur.534..512B, 2016ApJ...818L..22A}\footnote{For binary evolution, the low metallicity requirement is less stringent in some models \citep[][and reference therein.]{2016Natur.534..512B,2016A&A...588A..50M}}.

In this paper, we explore the effects of a large scale, dipolar surface magnetic field in suppressing wind mass loss and enabling an additional channel for a heavy BH to form in a solar metallicity environment.

In the last decade, large magnetometric surveys \citep{2015A&A...582A..45F, 2016MNRAS.456....2W} have revealed a population of magnetic massive stars, comprising $\sim10$\% of all main sequence OB stars. 
These magnetic fields, ranging from a few hundred gauss to tens of kilogauss, have properties different than dynamo-powered solar-type stars: they are large-scale and mainly dipolar, stable, and probably of fossil origin, i.e. they were left behind from a previous evolutionary epoch.

An important aspect of magnetic massive stars is the formation of wind-fed circumstellar magnetospheres \citep{2002ApJ...576..413U,2007MNRAS.382..139T,2008MNRAS.385...97U,2013MNRAS.428.2723U}. 
The interaction between the wind and field creates a region of closed loops (Fig 1) that channels the upflowing wind material into standing shocks near the loop apices. 
The magnetic field strongly couples the wind to the stellar surface, forcing it into co-rotation.
In absence of significant stellar rotation able to provide centrifugal support to the cooling, post-shock material, the trapped gas is pulled back to the stellar surface by gravity over a dynamical timescale \citep{2016MNRAS.462.3830O}. Such a magnetosphere is referred to as a ``dynamical magnetosphere'' (DM). The mass-loss rate is thus reduced according to the fraction of the stellar surface feeding closed loops.

It has been shown both theoretically and observationally that the rotational braking produced by these magnetic fields is very effective for the most massive O-type stars \citep{2009MNRAS.392.1022U,2013MNRAS.429..398P}\footnote{With the exception of Plaskett's star, which has significant rotation, and is thought to be a post mass-transfer object \citep{2013MNRAS.428.1686G}.}. Therefore very massive magnetic stars should rapidly transition from hosting a rotationally-supported magnetosphere to a DM. 
As we will present, many of the known magnetic O stars have a significant fraction of their winds returning to the stellar surface because of magnetic confinement, effectively reducing the mass-loss to a point that can rival with the effect of a low metallicity.

In this paper, we explore how the magnetic confinement evolves with time to predict how large-scale, dipolar magnetic fields, like those measured on $\sim$10\% of O stars, will reduce the lifetime-integrated mass-loss, making it easier to form heavy BHs from magnetic progenitors, lessening (or altogether doing away with) the requirements for very low metallicity.

\begin{figure}
\includegraphics[width=0.49\textwidth]{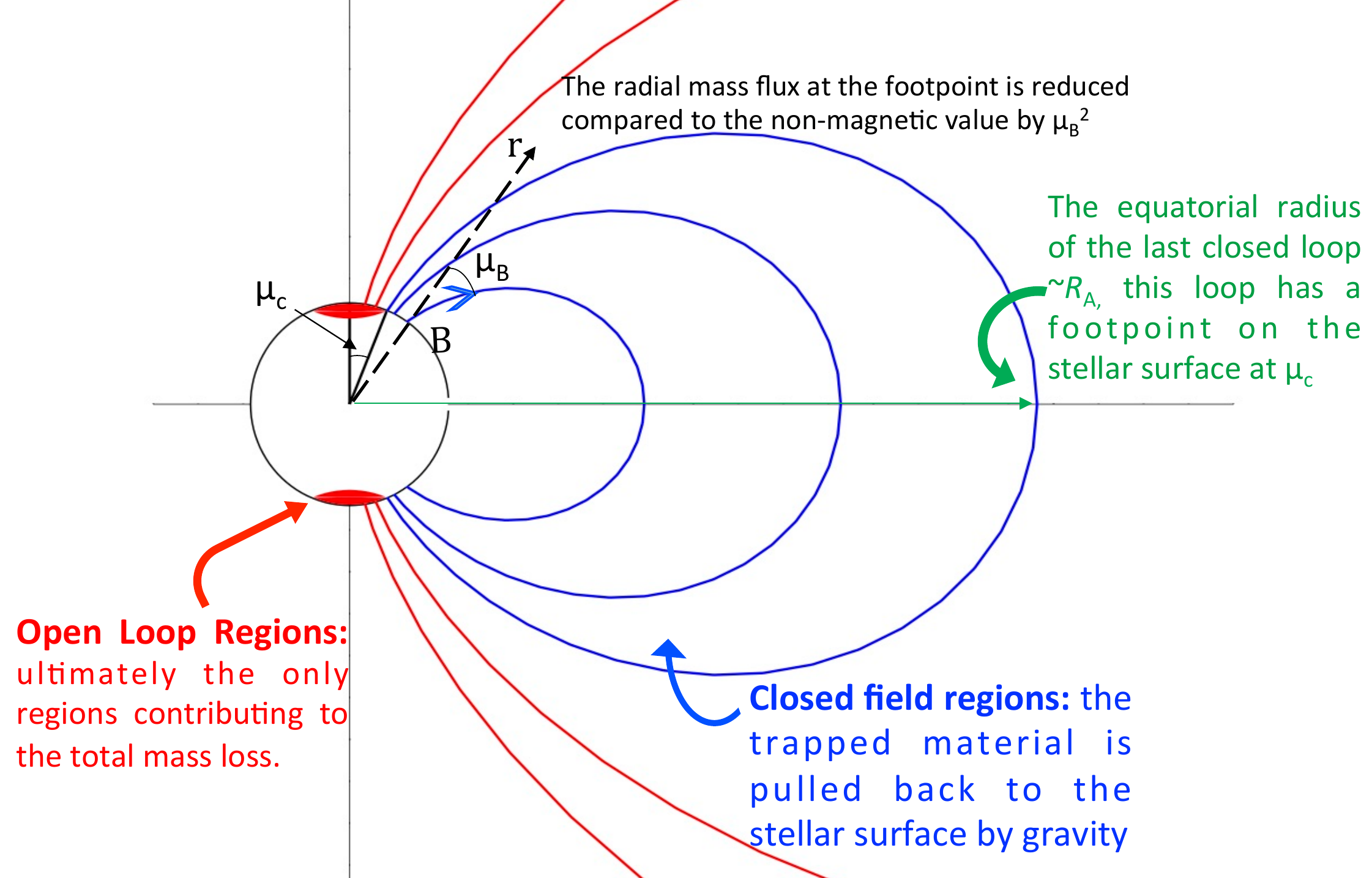}
\caption{\label{fig:cartoon} Schematic representation of the circumstellar magnetosphere of a slowly-rotating magnetic massive star, based on the description of \citet{2002ApJ...576..413U,2008MNRAS.385...97U}. The equatorial radius of the last closed loop is given by the closure radius $R_c$, which is on the order of the Alfv\'en radius $R_A$ where the magnetic energy density balances the wind kinetic energy density. The whole magnetospheric structure co-rotates with the stellar surface.}
\end{figure}

Section \ref{sec:confine} summarises the current day magnetic confinement of magnetic O-type stars and compares with metallicity relations. 
Section \ref{sec:mesa} explains our implementation of the magnetic confinement scenario within the MESA evolution code. 
Section \ref{sec:results} presents the relation between the initial and final mass of our models at galactic metallicity. 
Finally, Section \ref{sec:conclusion} summarises our findings.

\section{Wind quenching by magnetic confinement in O-type stars}
\label{sec:confine}

As described above, a large-scale magnetic field at the surface of a massive star can confine the outflowing, radiatively-driven wind \citep{1997ApJ...485L..29B,1997A&A...323..121B}. The principal influence of the magnetic field on the stellar wind is to reduce the effective rate of mass loss, due to two main effects:

(i) In the slowly-rotating, dynamical magnetosphere case, only the open field regions contributes to the total mass loss by the star (red regions in Fig \ref{fig:cartoon}), as the trapped, post shock material located in closed-line regions (blue region in the figure) is constantly pulled back to the stellar surface by gravity \citep{2008MNRAS.385...97U}. 

(ii) The tilt of the magnetic field with respect to the direction normal to the stellar surface reduces the wind-feeding rate at the loop footpoint \citep{2004ApJ...600.1004O, 2016MNRAS.462.3672B}. This results in a further reduction of the total wind feeding rate, more important for low latitude loops near the magnetic equator. As low latitude field loops will generally be closed for a dipolar magnetic geometry with a typical wind confinement, this effect adds only marginally to the reduction of the mass loss rate; hence we ignore this higher order effect. However, as a consequence we will obtain a conservative lower limit to the mass-loss reduction caused by the presence of the magnetic field. 

According to \citet{2002ApJ...576..413U}, the equatorial radius, $R_c$, of the farthest closed magnetic loop in a magnetized wind with a dipolar geometry at the stellar surface is of the order of the Alfv\'en radius $R_A$ (see Fig. \ref{fig:cartoon}).  More precisely,
\begin{equation}
\label{eq:rc}
R_c \approx R_\star + 0.7(R_A - R_\star),
\end{equation}
where $R_\star$ is the stellar radius. 

The location of the Alfv\'en radius corresponds to the point in the magnetic equatorial plane where the field energy density equals the wind kinetic energy,
    \begin{equation}
    \label{eq:ra}
    \frac{R_A}{R_*} \approx 0.3 + \left(\frac{B_{p}^{2}R_\star^{2}}{4\dot{M}_{B=0}V_{\infty, B=0}} +0.25\right)^{1/4},
    \end{equation}
where $B_p$ is the surface dipolar field strength. 
It is important to note that the Alfv\'en radius is parametrised by the mass-loss rate and wind terminal velocity the star would have in absence of a magnetic field, $\dot{M}_{B=0}$ and $V_{\infty, B=0}$. This mass-loss will be referred to as the ``wind-feeding rate'' at the base of the magnetosphere, to avoid confusion with the greatly reduced total rate of mass loss.

Tracing back the last closed loop to its footprint on the stellar surface, we can determine the fractional area covered by open field lines (red-shaded region in Fig \ref{fig:cartoon}) as a function of the closure radius of the last magnetic loop. Following \citet{2008MNRAS.385...97U}, we assume that this fraction of the surface alone (reproduced at both magnetic poles) is responsible for the total mass loss from the star and we define the dipolar escaping wind fraction $f_B$:

 \begin{equation}
    \label{eq:f}
    f_B = \frac{\dot{M}}{\dot{M}_{B=0}} = 1 - \sqrt{1-\frac{R_*}{R_c}}.
\end{equation}    

We note that in the case of a star with dynamically significant rotation, fallback occurs only for magnetic loops for which material is not centrifugally supported, i.e. with equatorial radii less than the Kepler co-rotation radius $R_K$ \citep[]{2008MNRAS.385...97U, 2007MNRAS.382..139T}. Without any loss of generality, $R_c$ in the above equation could be replaced with $R_K$ in such cases.

\begin{table*}
\caption{ \label{tab:ostars} List of known magnetic massive O-type stars with their spectral type, rotational period $P$, fiducial rotational period that would be needed for dynamically important rotation $P_\mathrm{CM}$, mass, dipolar field strength $B_p$, Alfv\'en radius $R_A$, and escaping wind fraction $f_B$. Columns 1-3 and 5-8 are reproduced from \citet{2013MNRAS.429..398P}. }
\centering
    \begin{tabular}{l   c   c c c c  c c   c}
    \hline
    \multicolumn{1}{c}{Star}  & Spec. Type & $P$ &  $P_\mathrm{CM}$ & $M_\star$ & $B_\mathrm{pole}$ & $R_K/R_*$ & $R_A$/$R_*$ & $f_B$ \\
    & & (d) &(d) & ($M_\odot$) & (kG) & & & (percent) \\
     \multicolumn{1}{c}{(1)} & (2) & (3) & (4) & (5) &  (6) & (7) &(8) &(9) \\
    \hline
    HD 148937  & O6f?p 			&7		& 2.1  &60	&1.0		& 4.3  & 1.8 	& 33 \\
    CPD -282561  & O6.5f?p 		&70		& 5.4  &43	&$>$1.7	& 19 & $>$3.1 	&$<$ 18 \\
    HD 37022 & O7Vp 			&15		&1.9  &45	&1.1		& 9.4 & 2.4 	& 24 \\
    HD 191612 &  O6f?p-O8fp 		&537		&8.2  &30	&2.5		& 57 & 3.7 	& 15 \\
    NGC 1624-2 &  O6.5f?cp-O8f?cp &158		&23    &34	&$>$20	& 41 & $>$11 	&$<$ 4 \\
    HD 47129 &  O7.5III 			& *		&6.8  &56	&$>$2.8	& $<$ 2.2 & $>$5.4 	&$<$ 24  \\
    HD 108 &  O8f?p 				&18000	&3.2  &43	&$>$0.50	& 526 & $>$1.7 	&$<$ 36 \\
    ALS 15218  & O8.5V 			&		&4.1  &28	&$>$1.5	&  & $>$3.6 	& $<$15 \\
    HD 57682 &  O9V 			&64		&3.8  &17	&1.7		& 24 & 3.7 	& 15 \\
    HD 37742 & O9.5Ib 			&7		&2.7  &40	&0.06	& 2.1 & 1.1 	& 70 \\
   \hline
    \end{tabular}\\
    $^*$ $v\sin i$ is measured to be $\sim$ 300 km\,s$^{-1}$ resulting in $P_\mathrm{rot}/ \sin i =1.8$ d \citep{2013MNRAS.428.1686G}.

\end{table*}

\begin{figure}
\includegraphics[width=0.48\textwidth]{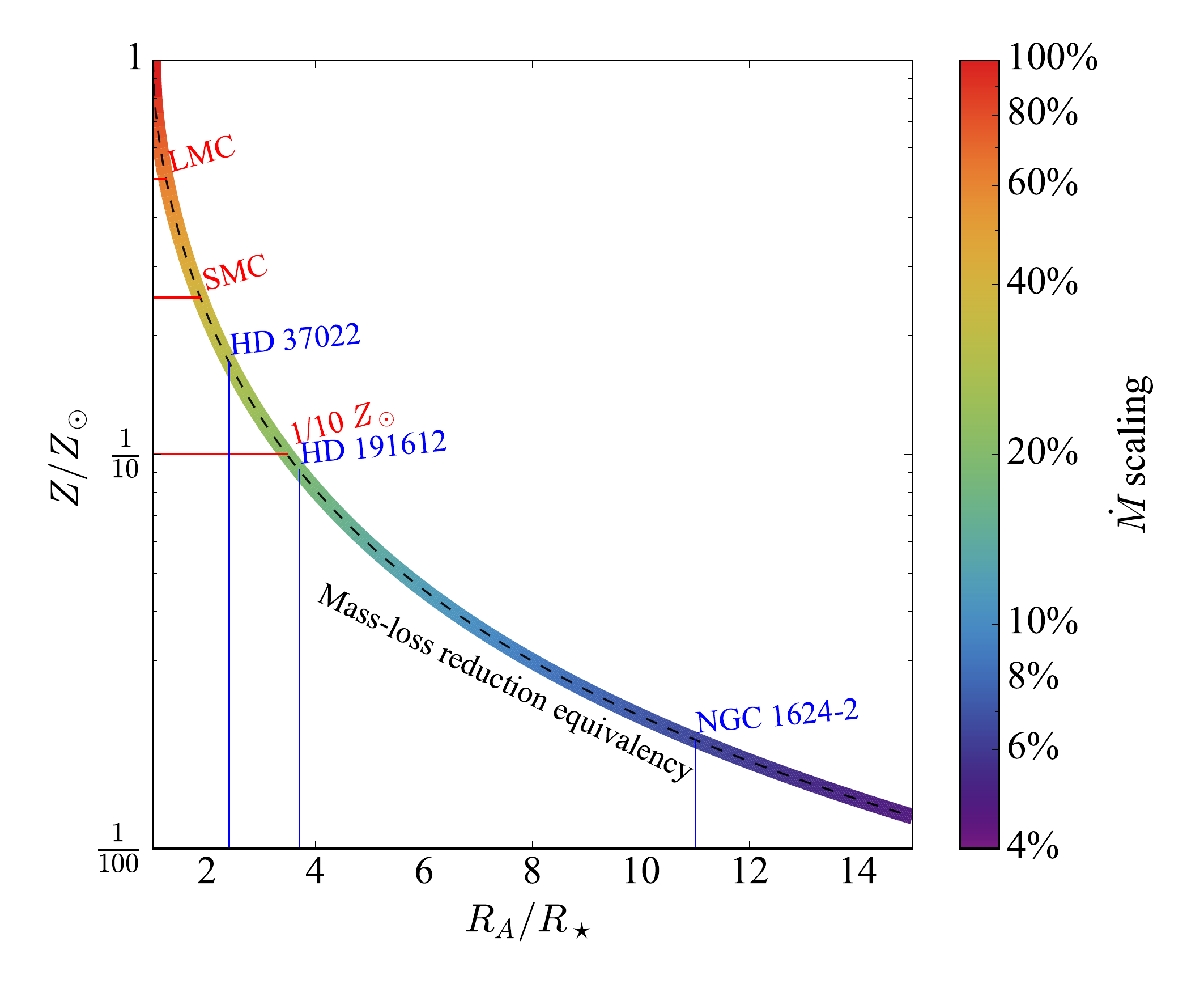}
\caption{\label{fig:metal} Equivalency curve between the reduction of mass-loss due to metallicity (in units of $\,Z_{\odot} = 0.019$) and the reduction of mass-loss due to magnetic wind confinement (expressed as the extent of the Alfv\'en radius). The curve is coloured according to the mass-loss scaling. The metallicity of the LMC and the SMC, as well as the $\sim$1/10$\,Z_{\odot}$ needed by single non-magnetic models to form heavy black holes \citep{2016ApJ...818L..22A}, are indicated by horizontal red lines. The Alv\'en radius of a few known magnetic O-type stars are indicated with vertical blue lines.}
\end{figure}

In Table 1, we compute the present-day values of $f_B$ for the known magnetic O-type stars included in the compilation of \citet{2013MNRAS.429..398P}. 
For all but one star the Alfv\'en radius is smaller than the co-rotation radius, as can be seen from columns 7 and 8, and we therefore use $R_A$ for our calculations. As a more intuitive comparison, we list in column 4 the fiducial rotational period that would be needed for the Kepler radius to be smaller than the Alfv\'en radius. For a generic magnetic main sequence O-type stars with $R_A\sim$2-3 $R_\star$ the rotational period would need to be shorter than one week in order to be dynamically significant, whereas the observed rotational periods are typically of the order of months. 

The typical known magnetic O-type stars have an Alfv\'en radius of $R_A$ $\simeq$ 1.1 - 3.7 $R_\star$ which corresponds to an escaping wind fraction $f_B$ of 70 -15 percent, respectively. 
NGC1624-2, the magnetic O-type star with the strongest field known \citep{2012MNRAS.425.1278W} has a much larger Alfv\'en radius, $R_A$ $\simeq$ 11 $R_\star$, leading to only 5 percent of its wind escaping the magnetosphere.
 
For comparison, Fig \ref{fig:metal} illustrates the metallicities and equivalent values of $R_A$ that produce an equivalent mass-loss reduction.  The mass-loss rate dependence on metallicity for non-magnetic stars is taken from the scaling relations by \citet{2001A&A...369..574V}. 

For most known magnetic stars in the Galaxy, the effect of the magnetic field corresponds to the equivalent mass-loss reduction for stars at metallicities ranging between that of the SMC and 1/10 $\,Z_\odot$. 
The most extreme case, NGC 1624-2, has a mass loss reduction equivalent to that which would occur for a similar star at a metallicity $\sim 1/30\ \,Z_\odot$. 
	
Naively, Galactic metallicity magnetic O stars could in principle evolve in a fashion similar to O stars with much lower metallicity. As a consequence, such evolution might then permit the formation of heavier remnants even at a metallicity higher than the $Z <  1/10\ \,Z_\odot$ required by non-magnetic, single star models for the formation of BH as massive as those involved in the merger of GW150914  \citep{2015MNRAS.451.4086S, 2016Natur.534..512B, 2016ApJ...818L..22A}. 	
	
However, given its dependence on magnetic, stellar, and wind parameters, the escaping wind fraction due to magnetic confinement will likely evolve during a star's lifetime in a different way than a reduction of mass-loss due to a low metallicity. 
Therefore as a first step, we concentrate on the total mass that is lost during the span of a star's main sequence lifetime, by implementing the effect of wind confinement in MESA, as described in the following section.

\vspace{1cm}

\section{The behaviour of magnetic confinement over stellar evolution time scales}
\label{sec:mesa}

The well-studied large-scale magnetic fields that have been firmly detected at the surfaces of many massive stars -- believed to have a fossil origin \citep[e.g.][]{2009MNRAS.397..763B,2011MNRAS.416.3160W} -- have only been considered thus far in a handful of evolutionary models. 

\citet{2011A&A...525L..11M} studied the effects of magnetic fields in enhancing surface angular momentum loss by magnetic braking during the evolution of a 10$\,M_\odot$ star.	Considering two models reflecting extreme behaviours of angular momentum transport in the interior, they found that (i) when the interior is differentially rotating, the surface becomes enriched in nitrogen (ii) when the interior is in solid-body rotation, there is no surface nitrogen enrichment and the surface rotation decreases more rapidly with time than for the differential rotation case. 

 Most other studies using magnetic stellar evolution models considered only the internal effects of shear dynamo-generated magnetic fields in massive stars \citep{2005ApJ...626..350H, 2005A&A...435..247P, 2003A&A...411..543M,2004A&A...422..225M,2005A&A...440.1041M}, mostly to account for the transport angular momentum via the proposed (but debated) Tayler-Spruit mechanism \cite[][see evolutionary models from, e.g., \citealp{2011A&A...530A.115B, 2016A&A...585A.120S}]{1973MNRAS.161..365T, 2002A&A...381..923S}.

 Such dynamo mechanisms have been proposed to operate in massive stars, either generated by convection in the core \citep{2016ApJ...829...92A} or in a subsurface layer \citep{2009A&A...499..279C}, and also generated by shear in the radiative envelope \cite[][but see \citealp{2007A&A...474..145Z}]{2002A&A...381..923S,2006A&A...449..451B, 2012MNRAS.425.2267R}. However, the presence of dynamo fields detectable at the surfaces of hot, massive stars have not been confirmed by observations yet \cite[e.g.][]{2015IAUS..305...61N}.

  While it is clear that fossil or possible dynamo fields do have an impact on the stellar interiors and especially angular momentum transport \citep[e.g.][]{2010A&A...517A..58D,2010MNRAS.402..271D, 2011A&A...525L..11M}, our immediate focus is to study how large-scale fossil fields affect the evolution of the mass loss, and ultimately on the final mass available to form a stellar remnant. 
   
 Thus, in terms of 1D hydrodynamical model calculations, we account for the alteration of the mass-loss rates due to magnetospheric effects for stars with large initial masses ($>40 \,M_\odot$, corresponding to O-type stars), as described below.   Furthermore, we consider only non-rotating models, to reduce the complications arising from the unknown structure of fossil fields in stellar interiors, and the associated modification of the interior angular momentum transport. This is a very reasonable first approach considering that most known magnetic O-type stars rotate very slowly compared to non-magnetic O-type stars (see the rotational periods summarised in Tab \ref{tab:ostars}). Details of our implementation are described in the following subsections.

\subsection{MESA implementation}

We use the open-source 1D hydrodynamical stellar evolution code, Modules for Experiments in Stellar Astrophysics \citep[MESA,][]{2011ApJS..192....3P,2013ApJS..208....4P,2015ApJS..220...15P} with the following new, simple treatment to manipulate the mass-loss rates.
		
We implement the effect of a large-scale dipolar magnetic field using the mass loss reduction prescription of \cite{2002ApJ...576..413U} as summarised in \S\ref{sec:confine}. We assume a range of initial (ZAMS) surface magnetic fluxes, corresponding approximately to the range of fluxes measured in real magnetic O stars. We impose flux conservation as the models are allowed to evolve; hence the surface magnetic field strength changes with time according to:
				
\begin{equation}
 F \sim 4\pi R_\star^2(t) B_p(t) = \mbox{constant}. \label{eq:flux}
 \end{equation}
The polar field strength at the surface will therefore scale as:
 \begin{equation}
    B_{p}(t) = B_{p,0} \left(\frac{R_{\star,0}}{R_\star(t)}\right)^2,
  \end{equation}
where $B_{p,0}$ and $R_{\star,0}$ correspond to the polar field and stellar radius defined at the start of the evolution. 

With the obtained polar field strength $B(t)$, the non-magnetic Vink ``wind-feeding'' rate\footnote{We note here that the wind properties are dependent on the adopted metallicity described in the following subsection.}, terminal velocity (see sec. \ref{sec:mdot}), and the radius, we calculate the Alfv\'en radius from equation \ref{eq:ra}. It is straightforward then to obtain the escaping wind fraction $f_B(t)$ from equations \ref{eq:rc} and \ref{eq:f}.

The final mass loss for that time step is obtained by scaling the current time step ``wind-feeding'' rate with the escaping wind fraction $f_B(t)$ allowing for mass to escape only via open loops, such that:

\begin{equation}
\dot{M}_\mathrm{final}(t) = f_B(t)  ~ \dot{M}_\mathrm{Vink}(t). \label{eq:fb}
\end{equation}

\subsection{Grid of models}

For consistency with the results discussed by \citet{2016ApJ...818L..22A}, we aim for simple model calculations that are comparable to the non-magnetic models presented by \citet{2010ApJ...714.1217B}.

\subsubsection{General properties of the models}

	 For our model calculations we adopt the hydrogen, helium and metal fractions as $X = 0.732$, $Y = 0.249$, $Z = 0.019$, and the chemical mixture of metals are from \citet[][the isotopic ratios are adopted from \citealp{2003ApJ...591.1220L}]{1989GeCoA..53..197A}. These values are the ones used in the original Vink prescription, as well as in the models presented by \citet{2010ApJ...714.1217B}.

The convective core boundary is determined by the Schwarzschild criterion, and we neglect overshooting. This can be justified by the large convective cores in this mass range \cite[see also][]{2015A&A...573A..71K}. We adopt a mixing length parameter $\alpha_{MLT} = 1.5$ \citep{2012sse..book.....K}. 

For this study, we follow our model calculations until core hydrogen exhaustion. This is reasonable since the presence and impact of fossil fields in the post-main sequence phases are very poorly understood. Due to the increase in the stellar radius after the TAMS, the fossil magnetic fields are expected to weaken significantly, while small-scale dynamo fields might take place due to the convective surface layers that develop at this phase of the evolution \citep{2009A&A...499..279C}. Therefore the wind confinement by fossil fields is expected to be small during the post-main sequence evolution, except in those stars with the very largest initial magnetic fields.
Another open question is the role played by \textit{interior} fossil magnetic field in the core collapse mechanism itself.

\subsubsection{Choice of initial masses}

Our choice of initial masses is motivated by our goal to explore the formation of heavy black holes at Galactic metallicity for magnetic stars of plausible initial masses. We note that the most massive magnetic O-type star known is $\sim 60\,M_\odot$ \citep[HD 148937;][]{2012MNRAS.419.2459W}. 
According to the models presented by \citet{2016Natur.534..512B} and \citet{2015MNRAS.451.4086S}, remnant of masses $>30 \,M_\odot$ could be formed by stars with initial masses ranging between $\sim$40-80$\,M_\odot$ for metallicities ranging between 0.01 and 0.5 $\,Z_\odot$.

We therefore consider a mass range using 40, 60 and 80 M$_{\odot}$ models. The 80M$_{\odot}$ models belong to the Very Massive Star (VMS) category \citep{2012ASPC..465..207V, 2015A&A...573A..71K}. These objects typically live very close to the Eddington limit, and can experience a variety of complex radiation-hydrodynamical instabilities \citep[see, e.g.][]{2015ApJ...813...74J}. Rather than simulating these instabilities in detail, for simplicity we adopt the \textsc{mlt++} prescription introduced by \citet{2013ApJS..208....4P}, which reduces the superadiabaticity in radiation-dominated convection zones and thereby allows models to be evolved successfully through the near-Eddington stages.

\subsubsection{Choice of initial magnetic field values}

 In our grid of models we adopt three realistic magnetic cases and a non-magnetic case. 
		As we are considering the flux conservation hypothesis, we compute our magnetic models using a set of three magnetic fluxes, defined as in equation \ref{eq:flux}. 
This means that for each magnetic strength group, the initial dipolar field strength at the ZAMS will be different for each initial mass, as the more massive stars have a larger ZAMS radius.

Known magnetic O-type stars generally have a radius of $\sim10R_\odot$ and a magnetic field of 1-2 kG \citep{2013MNRAS.429..398P}. Therefore the magnetic flux is of the order of $10^{28}$ G cm$^2$. In the case of the most magnetic O-type star NGC 1624-2, the magnetic flux reaches $10^{29}$ G cm$^2$ due to the higher dipolar strength \citep{2012MNRAS.425.1278W}. In the case of the supergiant HD\,37742, the larger radius and weaker dipole strength only lead to $10^{27}$ G cm$^2$ \citep{2015A&A...582A.110B}.
		
We therefore use magnetic fluxes of $10^{27}$, $10^{28}$, and $10^{29}$ G cm$^2$, respectively. The corresponding initial dipolar field strengths at the ZAMS will be presented in Fig. \ref{fig:Bpole} and discussed further in \S \ref{sec:results}.

\subsection{Mass-loss prescription}
\label{sec:mdot}

A very sensitive question is the treatment of mass-loss rates, and we note here that our purpose is to complement an existing mass-loss scheme with the effects of wind quenching from magnetic confinement. Therefore, while our results quantitatively depend on the adopted scheme, the qualitative influence is independent of the adopted wind description.  
We adopt the widely-used Vink rates \citep{2000A&A...362..295V,2001A&A...369..574V} for consistency reasons, and we did not manipulate the ``original'' MESA routine besides complementing it with a function accounting for the time-dependent reduction of the mass loss as described by equations \ref{eq:flux} to \ref{eq:fb}. However, we note here that recent studies indicate that there may be discrepancies between the theoretical Vink rates and mass-loss rates derived from state-of-the-art diagnostics.

(i) X-ray \citep{2014MNRAS.439..908C, 2013ApJ...770...80L, 2013A&A...551A..83H}, UV \citep{2013MNRAS.428.1837S, 2012arXiv1205.3075B, 2013A&A...559A.130S}, and IR \citep{2011A&A...535A..32N} diagnostics of massive stars are consistent with Vink rates reduced by a factor 2 when up-to-date abundances are considered \citep{2010A&A...512L...7V, 2016MNRAS.458.1999P}. 

(ii) Furthermore, the theoretical position of the first bi-stability jump \citep{1990A&A...237..409P, 1999A&A...350..181V, 2000A&A...362..295V}  has very recently been re-investigated and found to be at lower effective temperatures \citep{2016MNRAS.458.1999P}, while a large jump in mass-loss rates at the bi-stability is still debated \citep{2006A&A...446..279C, 2008A&A...478..823M}, and likely overestimated in evolutionary calculations \citep{2016arXiv161004812K}.

We note that we explicitly calculate the terminal velocity in equation~(\ref{eq:rc}) from the escape velocity $v_{\rm esc}$, adopting for $v_{\infty}/v_{\rm esc} = 2.6$ and $v_{\infty}/v_{\rm esc} = 1.3$ for the hot and cool side of the bi-stability jump, respectively \citep[][but see also \citealp{1998ASPC..131..218P,2006A&A...446..279C,2008A&A...478..823M,2010MNRAS.404.1306F,2012A&A...542A..79C}]{1995ApJ...455..269L, 2000ARA&A..38..613K}.

We also note here, that although some of our models may become LBVs during their main sequence, and thus enhanced mass-loss rates may need to be considered \citep[e.g.][]{2014A&A...564A..30G}, we do not account for this transition since we only aim at demonstrating how magnetic winds compare to a reference non-magnetic model.

\section{Results}
\label{sec:results}

\begin{figure*}
\includegraphics[width=0.99\textwidth]{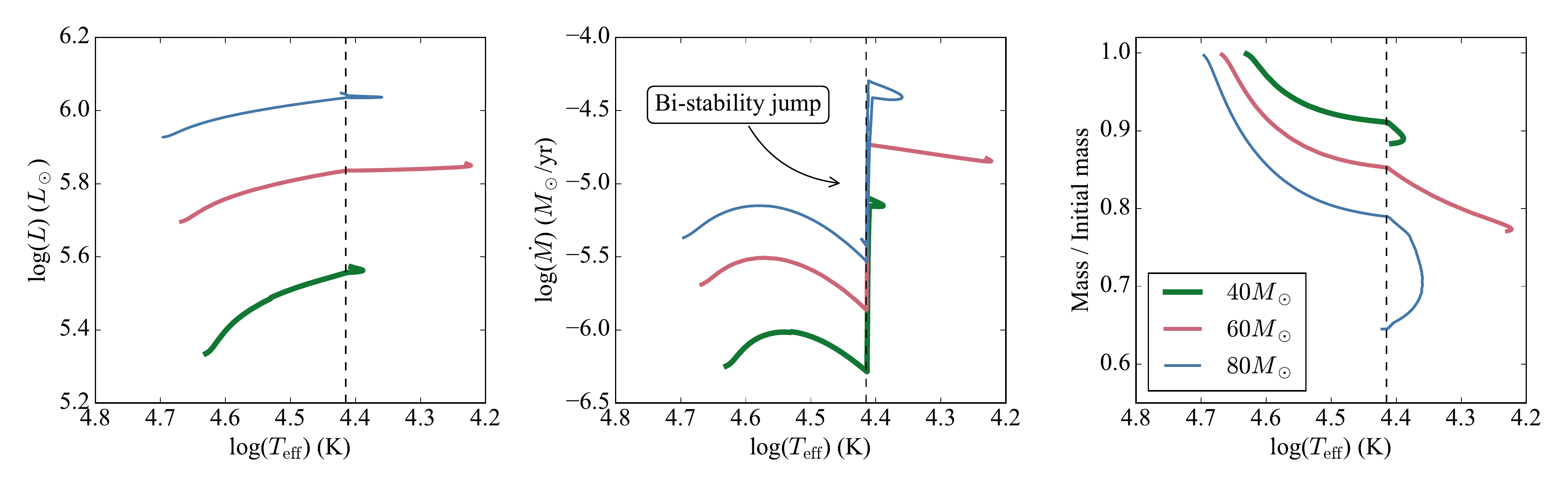}
\caption{\label{fig:noB} Left: evolutionary tracks of the non-magnetic models in the HR diagram. Middle: Wind mass-loss rate as a function of effective temperature using the Vink prescription as implemented in MESA. Right: Evolution of the stellar mass as a function of effective temperature, expressed as a fraction of the initial mass.  
The dashed vertical lines mark the change in trajectory in the HRD caused by the increase in mass-loss rate at the bi-stability jump, and therefore a more rapid decrease of the mass as a function of temperature. }
\end{figure*}

\begin{figure}
\includegraphics[width=0.48\textwidth]{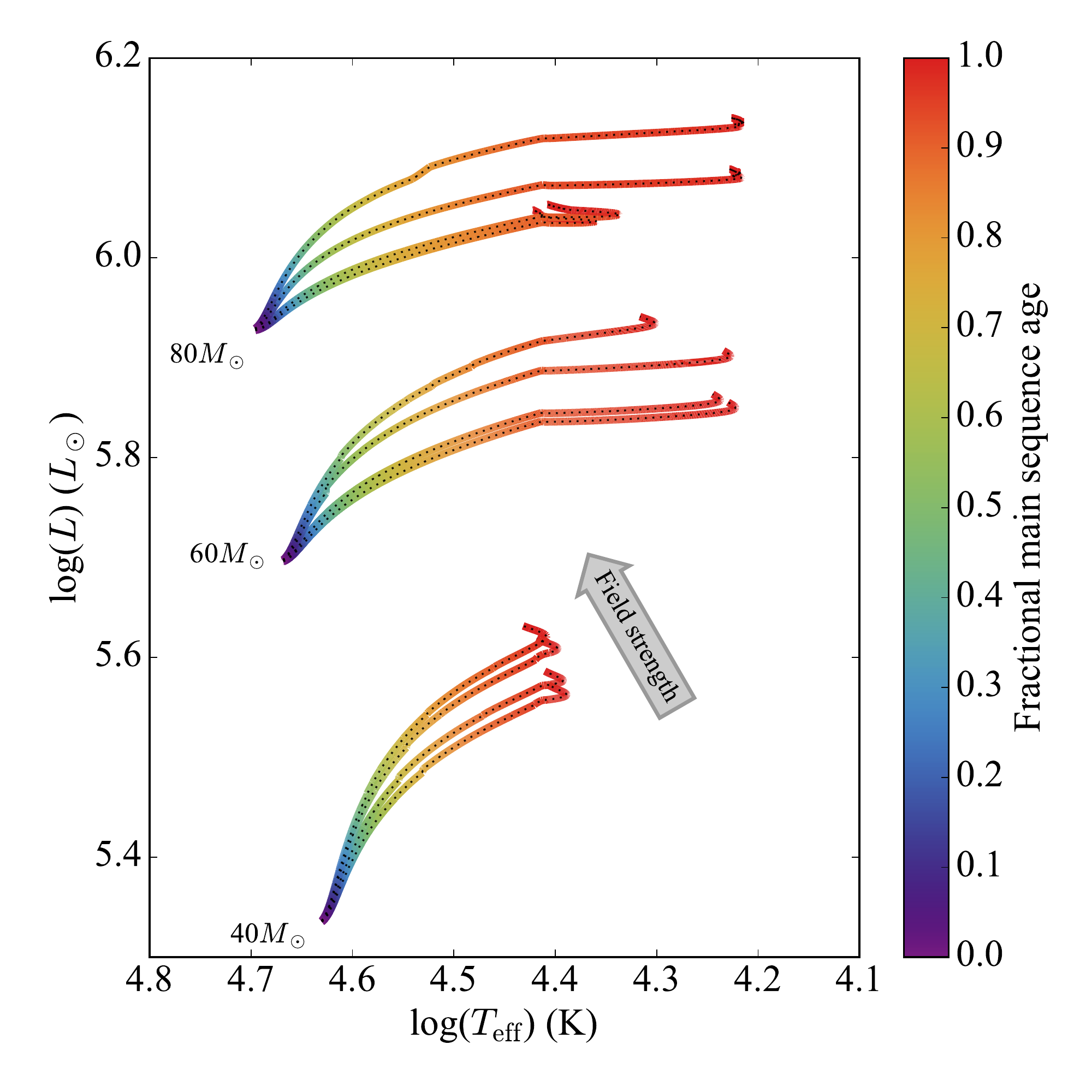}
\caption{\label{fig:HRD} Evolutionary tracks of all the models in the HR diagram. 
In each group, the lowest curve corresponds to the nonmagnetic model. The curves are coloured according to their fractional main sequence age, where 0 represent the ZAMS, and 1 represent the TAMS.  }
\end{figure}

\begin{figure}
\includegraphics[width=0.48\textwidth]{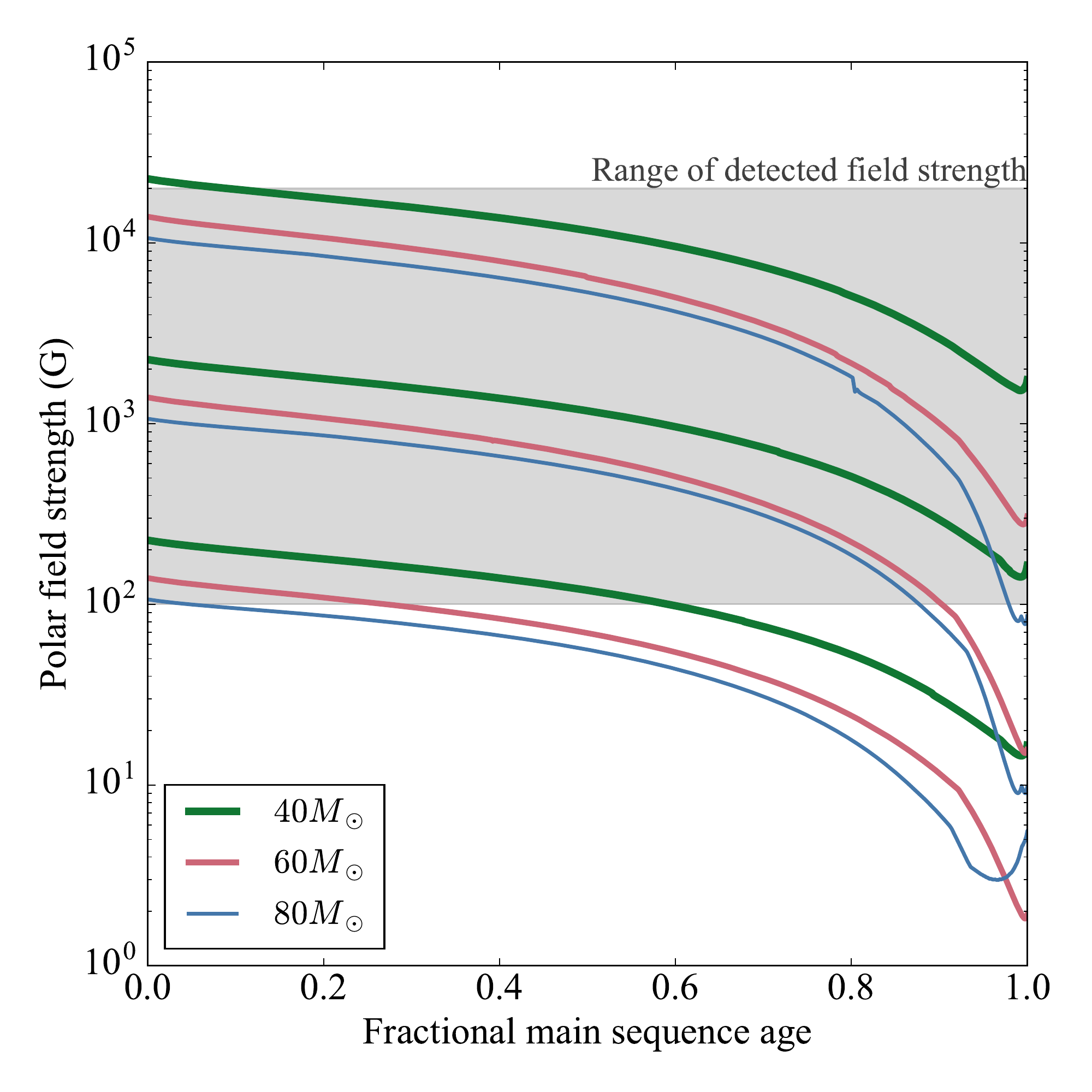}
\caption{\label{fig:Bpole} Evolution of the surface dipolar field strength as a function of fractional main sequence age. The grey zone illustrates the range of measured field strengths for known magnetic O-type stars. Each vertical group of curves correspond to a single value of magnetic flux -- the larger initial radius for larger masses leads to a lower initial surface field strength.}
\end{figure}

\begin{figure}
\includegraphics[width=0.48\textwidth]{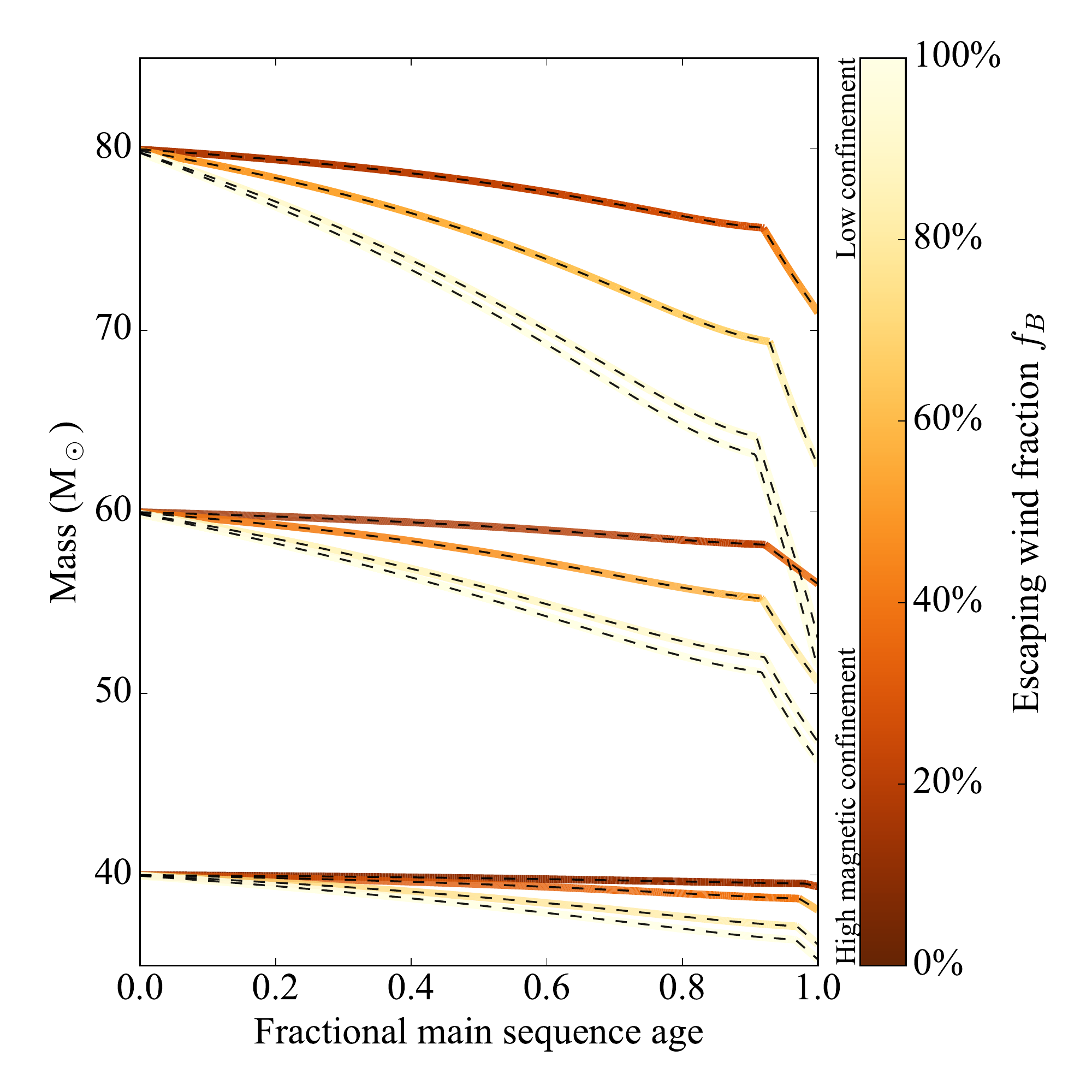}
\caption{\label{fig:mass} Stellar mass as a function of fractional main sequence age. For each initial mass group, the initial magnetic field values increase from bottom to top. The curves are coloured according to the escaping wind fraction $f_B$, to illustrate the regions of the parameter space where the magnetic confinement is important (lower $f_B$, darker colours).  }
\end{figure}

\begin{figure}
\includegraphics[width=0.48\textwidth]{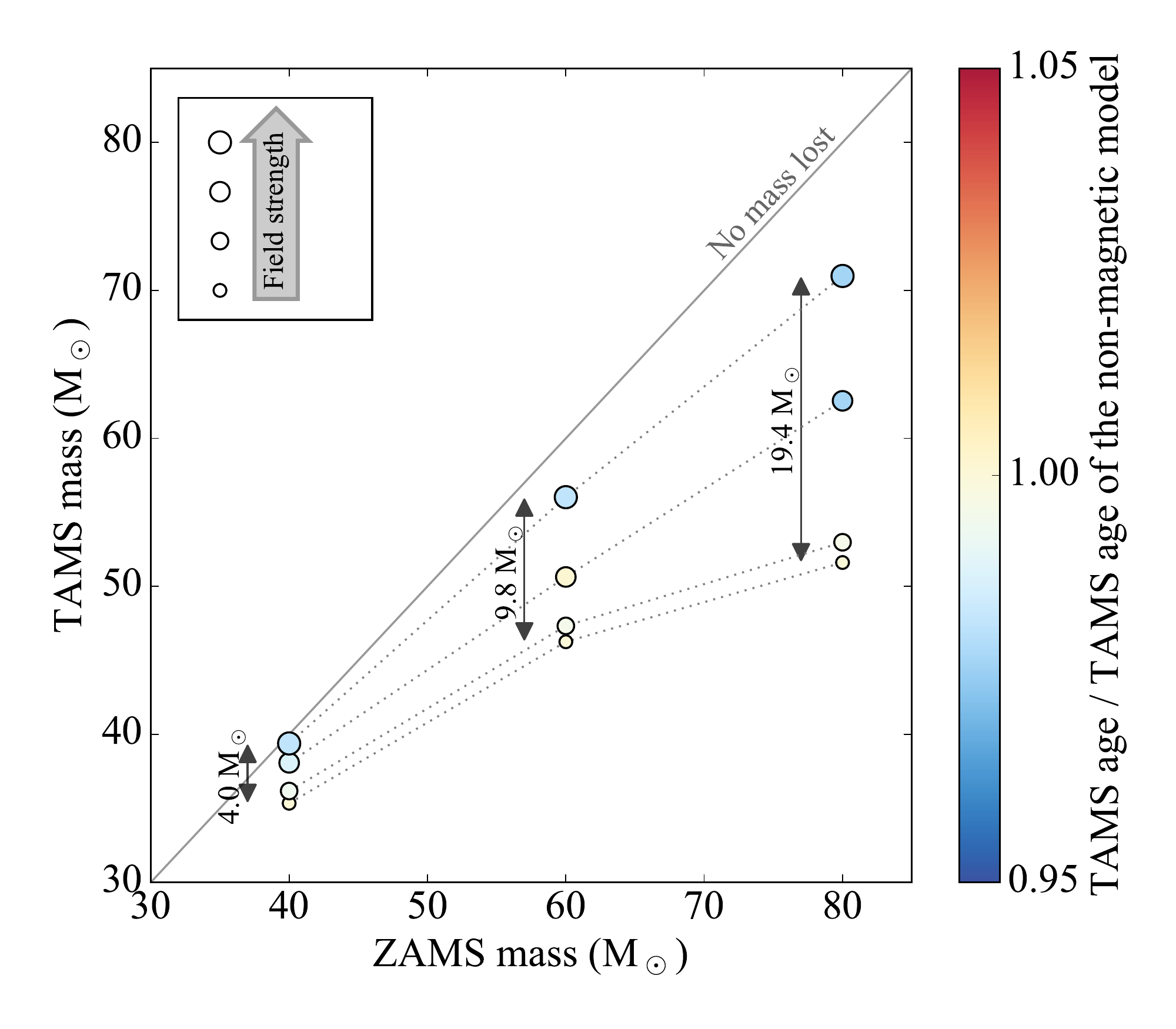}
\caption{\label{fig:final} Mass at the TAMS as a function of the initial mass at the ZAMS. The initial field strength increases with increased symbol size. The points are coloured according to the MS lifetime of the model as compared to the non-magnetic model of the same initial mass. The vertical arrows indicate the numerical value of the difference in TAMS stellar mass between the most magnetic and non-magnetic models. }
\end{figure}

\subsection{Non-magnetic models}

The left panel of Fig. \ref{fig:noB} shows the evolutionary tracks of the non-magnetic models for initial masses of $40\,M_\odot$, $60\,M_\odot$, and $80\,M_\odot$.
In most evolutionary tracks produced by our models, a change in trajectory is present around 26,000 K, as indicated by the vertical dashed line, after which the increase in luminosity with decreasing effective temperature is less steep.
The middle panel of Fig. \ref{fig:noB} shows the evolution of the mass-loss rate as a function of temperature, and demonstrates that this change in the evolution path is associated with the sudden increase in mass-loss rate caused by the bi-stability jump \citep[][Keszthelyi et al., in press]{2000A&A...362..295V,2001A&A...369..574V,2008A&A...478..823M, 2016MNRAS.458.1999P}.

The right panel of Fig. \ref{fig:noB} shows the evolution of the stellar mass as a function of effective temperature expressed as a fraction of the initial mass. The change in trajectory in the HRD results from the adaptation of the stellar structure to the more rapidly decreasing mass. 

We remark here that our main sequence non-magnetic models show good agreement with the models by \citet{2010ApJ...714.1217B} -- we find that our $40 \,M_{\odot}$ MESA model with $Z = 0.019$ has a TAMS mass of $35.3 \,M_{\odot}$, and the $40 \,M_{\odot}$ of \citeauthor{2010ApJ...714.1217B} model at $Z = 0.020$ has a TAMS mass of $34.6 \,M_{\odot}$ (K. Belczynski, priv. comm.).

\subsection{Model evolution in the HR diagram} 

Fig. \ref{fig:HRD} shows all our models in the HR diagram. The evolutionary tracks are coloured according to their fractional main sequence (MS) age\footnote{In our models, the hydrogen abundance in the core varies nearly linearly with the fractional MS age.}, where this fractional age is zero at the ZAMS and unity at the TAMS. 

For each initial mass group, the more magnetic models evolve at significantly higher luminosity. This can be understood by their higher mass at a given age, as will be presented below in Fig. \ref{fig:mass}. 
However as can be seen from the isochrones -- represented by constant colours in Fig. \ref{fig:HRD} -- for each initial mass group, stars of similar MS fractional age still have similar effective temperatures. 

For the models with an initial mass of 40$\,M_\odot$, the TAMS is located at similar effective temperature. In contrast, for the 60$\,M_\odot$ and 80$\,M_\odot$ the TAMS is located at different effective temperatures. However, in all cases, the more magnetic models reach the TAMS slightly quicker than the less magnetic ones, but with a difference between the MS lifetimes of less than 5\% (as will be shown in Fig. \ref{fig:final} below).

\subsection{Surface field evolution} 

Fig. \ref{fig:Bpole} shows the evolution of the dipolar field strength of the magnetic models as a function of the fractional MS age. We remind the reader that we compute our models on a grid of magnetic flux values which correspond to a desired range of surface magnetic field strengths. As a consequence of this approach, larger initial dipolar field strengths are obtained for the less massive models due to their smaller initial radii. 

The shaded area in Fig. \ref{fig:Bpole} corresponds to the range of dipolar field strength values for known magnetic O-type stars \citep[][and references therein; see Table \ref{tab:ostars}]{2013MNRAS.429..398P}, illustrating that our models span a realistic range of field strengths for their entire evolution. 
		As our lowest initial field value for each mass leads to an evolution very similar to the non-magnetic case, an extension of our grid to lower initial field values is not necessary.

The decrease in field strength in our model is directly tied to an increase of the stellar radius through magnetic flux conservation. We can see from the coherent decrease of all curves that the radius evolution is not a strong function of the initial field strength during the first 75 percent of the main sequence. 
		
Near the TAMS however, we can see that the increase in radius and decrease in field strength is a stronger function of both the initial field strength and the initial mass. 
This is consistent with the fact that, for a given initial mass, the more strongly magnetic stars evolve at generally higher luminosity and larger radii towards the end of the MS, due to their higher mass caused by less mass-loss.
This effect is more pronounced for the models with larger initial mass.
		
This result is very interesting considering the recent study by \citet{2016A&A...592A..84F}, who proposed that the seemingly young age of the magnetic massive star population could be the result of a decay of their surface magnetic fields that is more rapid than that obtained by a simple magnetic flux conservation model, especially for higher mass stars. 
Our preliminary results suggest that the inclusion of the change in stellar structure and evolutionary tracks for stars with large-scale magnetic fields might in part explain such a rapid decrease in surface field strength by a larger increase in stellar radius during the main sequence than would be expected from non-magnetic evolution models. However, self-consistent age determination with magnetic evolution tracks will be necessary to explore the magnetic flux conservation or decay hypotheses further.

\subsection{Mass evolution} 

Fig. \ref{fig:mass} shows the evolution of the stellar mass as a function of the fractional MS age. The curves are coloured as a function of the escaping wind fraction $f_B$, with darker colours corresponding to high  magnetic confinement and therefore a lower $f_B$. The break towards the end of the MS correspond to the increased mass-loss rate that occurs after the bi-stability jump.

For the lowest initial magnetic flux value considered (second to bottom curve of each mass group), magnetic confinement is unimportant throughout the MS evolution, as the escaping wind fraction is always greater than $\sim90$ percent (light colours). 
For the two higher initial magnetic flux values (top two curves of each mass group), typical for the majority of known magnetic O-type stars, the magnetic confinement is important (dark colours) for most of the main sequence. For the strongest initial magnetic flux value, which corresponds to the strongest magnetic O-type star known, the magnetic confinement is still important after the bi-stability jump, all the way to the TAMS. 	
The growing difference in stellar mass with age indicates that the mass evolution changes strongly as a function of initial magnetic field strength, especially for higher initial masses.

Fig. \ref{fig:final} summarises our results by showing the TAMS mass versus the ZAMS mass of our models. Increasing initial field strength is illustrated with increasing symbol size. 
From low to high initial mass, the difference in masses at the TAMS between the non-magnetic and the most magnetic models amount to 4$\,M_\odot$, 10$\,M_\odot$, and 20$\,M_\odot$, respectively. This leads to the TAMS mass of the most magnetic model being 11, 21, and 38 percent larger than the TAMS mass of the non-magnetic model, respectively. 
	
The points in Fig. \ref{fig:final} are coloured as a function of the total main sequence lifetime relative to the lifetime of the non-magnetic model. The small difference in age at the TAMS between models of the same initial mass group ($<5$ percent) illustrates that the difference in final mass is primarily due to the wind quenching from magnetic confinement, as opposed to a very different MS lifetime. 

Of course the final mass of a black hole produced by such a star will depend on subsequent post-main sequence evolution and core-collapse. But as pre-collapse progenitors of $40M_\odot$ or more should directly form black holes without a supernova \citep{1999ApJ...522..413F}, we nonetheless can conclude that according to our models, massive stars with a strong dipolar magnetic field will have a significant head start for the potential production of heavy stellar black holes at Galactic metallicity.

\section{Discussion and conclusions}
\label{sec:conclusion}

 In this paper, we set out to explore a new pathway for the formation of single, ``heavy'' ($>25\,M_\odot$) black holes -- with masses such as those involved in the LIGO event GW150914 -- through magnetic wind confinement. 
 Although massive stars with large-scale, strong magnetic fields only comprise $\sim10$ percent of the galactic OB star population, magnetic confinement is still an effective way to quench mass-loss at galactic metallicity. This is unlike the formation of heavy BH from single, non-magnetic stars that requires a mass-loss reduction through a low metallicity environment.  Our main conclusions are as follow:		

(i) We first evaluated the current-day escaping wind fraction, $f_B$, for known magnetic O-type stars with large-scale, dipolar field, which describe the fraction of the stellar wind that escapes through open field lines. 
For most magnetic O-type stars, $f_B$ is 10-30 percent. For the most extreme case, the O-type star NGC 1624-2, $f_B$ is only 5 percent. 
 These values correspond to an upper limit to $f_B$, as we only considered the change in mass loss due to the material that is trapped in closed magnetic loops and falls back to the stellar surface; and ignored the second order reduction of the wind-feeding rate at the base of the wind due to the tilt of the magnetic field lines with respect to the radial direction.
	
(ii) When comparing these values with the mass-loss rate reduction due to a reduced metallicity, we found that most magnetic stars in  our Galaxy have a reduction of their mass-loss equivalent to that found for stars with a metallicity between that of the SMC and 1/10 $\,Z_\odot$. Again for the most extreme case of NGC 1624-2, this corresponds to the mass-loss of a non-magnetic star with a metallicity of 1/30 $\,Z_\odot$. 
Therefore, we estimated that wind confinement by a realistic dipolar field would provide a mass-loss reduction of the same order as the low metallicity ($Z \sim 1/10 \,Z_\odot$) required by a BH formation scenario from single, non-magnetic stars.
	
(iii) As the strength of magnetic confinement is expected to change with time, due to evolutionary changes in surface magnetic and wind properties, we explored the integrated mass lost over the main sequence life time.
We computed non-rotating, solar metallicity evolution models in MESA using a mass loss calculation that is modified by the time-dependent escaping wind fraction for a magnetized wind with a dipolar geometry at the stellar surface.
We found that stars with higher initial magnetic strength evolved at higher luminosity, but at similar temperature, than their less/non-magnetic counterparts. 
The more magnetic models reach the TAMS slightly quicker than the less magnetic models, but with a difference between the MS lifetimes of less than 5 percent.

(iv) Our models show a significant difference in mass at the end of the main sequence for the most massive and most magnetic stars.
For the three initial masses considered (40, 60, and 80 $\,M_\odot$), the difference in mass at the TAMS between the non-magnetic and the most magnetic models (corresponding roughly to the magnetic flux of NGC 1624-2) amounts to 4$\,M_\odot$, 10$\,M_\odot$, and 20$\,M_\odot$, respectively.  This leads to the TAMS mass of the most magnetic model being 11, 21, and 38 percent larger than the TAMS mass of the non-magnetic model. 
Therefore according to our models, massive stars with a strong dipolar magnetic field will have a significant head start for the potential production of heavy stellar black holes at Galactic metallicity.

In our study, we imposed two main simplifications to our evolution models: 

(i) We imposed surface magnetic flux conservation as the models were allowed to evolve. 
There is some evidence that the measured magnetic field at the surface of massive stars might decrease more rapidly than what would be explained by the increase in stellar radius with age \citep{2016A&A...592A..84F}. 
A decay of the magnetic flux in time would modify the mass evolution of the magnetic model by enabling more mass-loss. 

(ii) We used non-rotating models under the assumption that the magnetic spin-down of massive O-type stars occurs very rapidly, as suggested by their generally very long rotation periods. According to the models of \citet{2011A&A...525L..11M}, however, the TAMS internal structure of a 10$\,M_\odot$ star that was born with slow rotation would be different than that of a similar star that was magnetically spun-down (assuming solid body rotation), the latter one having a larger core.  The effect of rotation in the evolution of massive magnetic O-type stars will be explored in a subsequent publication.

\section*{Acknowledgments}
The authors would like to thank K. Belczynski for providing the specifics of the models presented in \citet{2010ApJ...714.1217B,2016Natur.534..512B}  for comparison with ours. 
The authors would also like to thank the referee, Dr. Georges Meynet, for his useful comments and suggestions. 
VP acknowledges support provided by 
(i) the National Aeronautics and Space Administration through Chandra Award Number GO3-14017A issued by the Chandra X-ray Observatory Center, which is operated by the Smithsonian Astrophysical Observatory for and on behalf of the National Aeronautics Space Administration under contract NAS8-03060.
(ii) program HST-GO-13734.011-A that was provided by NASA through a grant from the Space Telescope Science Institute, which is operated by the Association of Universities for Research in Astronomy, Inc., under NASA contract NAS 5-26555.

DHC acknowledges support from Chandra Award Number TM4-15001B.

RHDT acknowledges support from NSF SI$^{2}$ grant ACI-1339600 and NASA TCAN grant NNX14AB55G.

GAW acknowledges Discovery Grant support from the Natural Sciences and Engineering Research Council (NSERC) of Canada

AuD acknowledges support by NASA through Chandra Award numbers GO5- 16005X, AR6-17002C, G06-17007B and proposal 18200020 issued by the Chandra X-ray Observatory Center which is operated by the Smithsonian Astrophysical Observatory for and behalf of NASA under contract NAS8- 03060.

\bibliographystyle{mn2e_fix2}
\bibliography{database}


\label{lastpage}
\end{document}